\documentstyle[11pt,moriond,epsfig]{article}

\def\ra{\rightarrow}

\def\be{\begin{equation}}
\def\ee{\end{equation}}
\def\bea{\begin{eqnarray}}
\def\eea{\end{eqnarray}}

\def\ra{\rightarrow}

\def\bea{\begin{eqnarray*}}
\def\ena{\end{eqnarray*}}
\def\beq{\begin{equation*}}
\def\enq{\end{equation*}}

\def\TT{\textstyle}
\def\D0{D\O~}

\begin{document}








{\small
\noindent
{hep-ph/9807316} \hfill {CTEQ-851}\\
{May 1998} \hfill {MSUHEP-80531}
}
\vspace*{1cm}

\title{GLUON RESUMMATION IN VECTOR BOSON PRODUCTION AND DECAY\footnote{
Talk given at the XXXIIIrd Rencontres de Moriond, 
``QCD and High Energy Hadronic Interactions'', 21-28 March 1998,
Les Arcs, France.}
}
\author{ C.-P. YUAN }

\address{Department of Physics and Astronomy, 
Michigan State University, \\
East Lansing, MI 48824, USA}

\maketitle\abstracts{
After a brief review on the technique of resumming the large 
logarithmic terms $\alpha^n_s \ln^m(Q^2/Q^2_T)$ due to the effects of 
multiple gluon emission predicted by the QCD theory,  
I discuss its application in the production of $W^\pm$, $Z$
and $\gamma \gamma$ in hadron collision.}

\section{Introduction}

It is the prediction of the QCD theory that at hadron colliders the 
production of Drell-Yan pairs or weak gauge bosons ($W^\pm$ and $Z$)
are often accompanied by gluon radiation.
Therefore, to test the QCD theory or to probe the electroweak properties
in the vector boson productions, it is necessary to include the 
effects of multiple gluon emission.
At the Tevatron (a $p \bar p$ collider), we expect 
about $2\times 10^{6}$ $W^\pm$ and $6 \times 10^{5}$ $Z$ bosons
produced at $\sqrt{S}=1.8$\,TeV, per $100\,{\rm pb}^{-1}$ of 
luminosity. This large sample of data is useful for 
(i) QCD studies (single and multiple scale cases), 
(ii) precision measurement of the $W$ boson mass and width, 
and
(iii) probing for new physics (e.g., $Z'$), {\it etc}.
To achieve the above physics goals requires detailed information on 
the distributions of the rapidity and the transverse momentum 
of $W^\pm/Z$ bosons and of their decay products.

Consider the production process $h_1 h_2 \rightarrow V X$.
Denote $Q_T$ and $Q$ to be the transverse momentum and the invariant
mass of the vector boson $V$, respectively. 
When $Q_T \sim Q$, there is only one hard scale in this problem,
and the fixed-order perturbation calculation is reliable.
When $Q_T \ll Q$, this becomes a two scale problem, and 
the convergence of the conventional
perturbative expansion becomes impaired.
Hence, it is necessary to apply the technique of QCD resummation to
resum the {\it singular} terms:
\begin{eqnarray*}
\frac{d\sigma}{dQ_T^2} \sim \frac{1}{Q_T^2}  
\left\{ \right.
      \alpha_S (L + 1) &+ \alpha_S^2 (L^3 + L^2) 
&+ \alpha_S^3 (L^5 + L^4) +
      \alpha_S^4 (L^7 + L^6) + ... \\ 
                         &+ \alpha_S^2 (~L + ~~1) 
&+ \alpha_S^3 (L^3 + L^2) +
      \alpha_S^4 (L^5 + L^4) + ... \\
                   &    &+ \alpha_S^3 (~L + ~~1) 
+ \alpha_S^4 (L^3 + L^2) + ...\left. \right\},
\end{eqnarray*}
where $L$ denotes $\ln(Q^2/Q_T^2)$ and the explicit coefficients 
multiplying the logs are suppressed.

Resummation of large logarithms yields a Sudakov form factor 
and cures divergence at $Q_T \ra 0$.
It was pioneered by Dokshitzer, D'yakonov and Troyan (DDT) 
who performed the analysis in $Q_T$-space which lead to the
leading-log resummation formalism \cite{DDT}.
Later, Parisi-Petronzio showed \cite{Parisi} that for 
large $Q$ the $Q_T \ra 0$ region can be calculated perturbatively by 
imposing the condition of the transverse momentum conservation,
\bea
\delta^{(2)} \left( \sum_{i=1}^{n} {\vec k_{T_i}} - {\vec q_T} \right)
=\int {d^2b \over 4 \pi^2} \;\TT{e}^{i{\vec{Q}_T}\cdot{\vec b}}
\prod_{i=1}^{n} \;\TT{e}^{i{\vec k_{T_i}}\cdot{\vec b}},
\ena
in the $b$-space (the impact parameter, the Fourier conjugate to $Q_T$).
This improved formalism sums also some subleading-logs.
As $Q \ra \infty$,
events at $Q_T \sim 0$ may be obtained asymptotically  by the 
emission of  at least two gluons whose transverse momenta are 
not small and add to zero.
The intercept at $Q_T=0$ is predicted to be \cite{Parisi}
\bea
\left.\frac{d\sigma }{dQ_T^2}\right| _{Q_T \ra 0} \sim 
\sigma_0 \left( {\Lambda^2 \over Q^2} \right)^{\eta_0},
\ena
\noindent
where $\eta_0=A\ln\left[1+{1\over A}\right]$
with $A=12 C_F/(33-2 n_f)$, and
$\eta_0\simeq 0.6$ for $n_f=4$ and $C_F=4/3$. 
Collins-Soper-Sterman (CSS) extended \cite{CSS} the work by 
Parisi-Petronzio in $b$-space 
and applied the properties of the renormalization group invariance 
to set up a formalism that resums all the large log terms 
to all orders in $\alpha_s$.
This is the formalism I will concentrate on in this talk.

Recently, there is another theory model in $Q_T$-space
(extension of DDT) proposed by Ellis and Veseli \cite{ev},
which does not have either 
the exact transverse momentum conservation
or the renormalization group invariance conditions.
The reader can find a detailed discussion of this formalism 
in the talk by K. Ellis at this conference. 
Despite of the imperfectness of the formalism, it can still be useful
for the $W^\pm$ and $Z$ physics program at the Tevatron.      
I shall come back to this point in the conclusion section.

\section{Collins-Soper-Sterman (CSS) resummation formalism}
\indent

In the resummation formalism by Collins, Soper and Sterman \cite{CSS}, 
the cross section is written in the form
\bea
{\frac{d\sigma (h_1h_2\rightarrow VX)}{dQ^2\,dQ_T^2dy\,\,}}=
{\frac 1{(2\pi)^2}}\int d^2b\,e^{i{\vec{Q}_T}\cdot {\vec{b}}}
{\widetilde{W}(b,Q,x_1,x_2)}+~Y(Q_T,Q,x_1,x_2),
\ena
where $Y$ is
the {\it regular} piece which can be obtained by subtracting the 
{\it singular} terms from the exact fixed-order result. 
${\widetilde{W}}$ satisfies a 
{ renormalization group equation}.
Its solution is of the form
\bea
{\widetilde{W}(Q,b,x_1,x_2)}=e^{-{\cal S}(Q,b{,C_1,C_2})}
{\widetilde{W}\left(\frac{C_1}{C_2b},b,x_1,x_2\right)},
\ena
where the Sudakov exponent is defined as 
\bea
{\cal S}(Q,b{,C_1,C_2})=\int_{C_1^2/b^2}^{C_2^2 Q^2}
\frac{d\overline{\mu }^2}{
\overline{\mu }^2}\left[ {A}\left( {\alpha}_s(\overline{\mu }),C_1\right) 
\ln
\left( \frac{C_2^2 Q^2}{\overline{\mu }^2}\right) 
{+B}\left( {\alpha}_s(\overline{\mu }),C_1,C_2\right) \right],
\ena
and the $x_1$ and $x_2$ dependence of $\widetilde{W}$ factorizes as
\bea
{\widetilde{W}\left(\frac{C_1}{C_2b},b,x_1,x_2\right)}=
\sum_je_j^2\;{\cal C}_{jh_1}\left(\frac{C_1}{C_2b},b,x_1\right)\;
{\cal C}_{jh_2}\left(\frac{C_1}{C_2b},b,x_2\right),
\ena
where ${\cal C}_{jh}$ is a convolution of the parton distribution with a
calculable Wilson coefficient, called $C_{ja}$ function: 
\bea
{\cal C}_{jh}(Q,b,x)=\sum_{a}\int_x^1{\frac{d\xi }\xi }\;
C_{ja}\left(\frac{x}{\xi},b,\mu={C_3 \over b},Q\right)\;
f_{a/h}\left(\xi ,\mu={C_3 \over b} \right),
\ena
where $a$ sums over incoming partons, and $j$ denotes the quark flavors 
with (electroweak) charge $e_j$.
A few comments about this formalism is listed below:
      \begin{itemize}
\item 
The $A$, $B$ and $C$ functions can be calculated order-by-order
in $\alpha_s$.
\item
A special choice of the renormalization constants $C_i$ 
can be made so that the singular terms obtained from the expansion 
of the CSS resummed calculation agrees with that from the fixed-order 
calculation. This is the canonical choice. It has
$C_1=C_3=2e^{-\gamma _E}\equiv b_0$ and $C_2=C_1/b_0=1$,
where $\gamma _E=0.577\dots$ is Euler's constant.
\item  
$b$ is integrated from 0 to $\infty$.
For  $b\gg 1/\Lambda_{QCD}$, the perturbative calculation is 
no longer reliable.
Hence, a non-perturbative function is needed in this formalism to 
compare theory prediction with experimental data.
      \end{itemize}
We refer the readers to Ref. \cite{wres} for a more detailed discussion on 
how to apply the CSS resummation formalism
to phenomenological physics at hadron colliders.

\subsection{Non-perturbative function}

As noted in the previous section, it is necessary to include a 
non-perturbative function in the CSS resummation formalism to 
incorporate some long distance physics that is not accounted by the
perturbative derivation.
Collins-Soper postulated \cite{CSS}
\bea
\widetilde{W}_{j{\bar{k}}}(b)=
\widetilde{W}_{j{\bar{k}}}(b_{*})\widetilde{W}
_{j{\bar{k}}}^{NP}(b)\,,
\ena
with
\bea
b_{*}={\frac b{\sqrt{1+(b/b_{max})^2}}}\,, 
\ena
so that $b$ never exceeds $b_{max}$ and  
$\widetilde{W}_{j{\bar{k}}}(b_{*})$ can be reliably calculated
perturbatively.
(In a numerical calculation, $b_{max}$ is set to be, say, 
$1/(2\,{\rm GeV})$.)
Based upon a renormalization group analysis, they found that
the non-perturbative function can be generally written as 
\bea
\widetilde{W}_{j\bar{k}}^{NP}(b,Q,Q_0,x_1,x_2)=\exp \left[ -F_1(b)\ln 
\left( \frac{Q^2}{Q_0^2}\right) -F_{j/{h_1}}(x_1,b)-F_{{\bar{k}}/{h_2}   
}(x_2,b)\right]\,, 
\ena
where $F_1$, $F_{j/{h_1}}$ and $F_{{\bar{k}}/{h_2}}$ must be
extracted from data with the constraint 
\bea
\widetilde{W}_{j\bar{k}}^{NP}(b=0)=1.
\ena
Furthermore,
$F_1$ only depends on $Q$, while 
$F_{j/{h_1}}$ and $F_{{\bar{k}}/{h_2}}$ in general depend on 
$x_1$ or $x_2$.
Later, in Ref. \cite{sterman}, it was shown that the 
$F_1(b)\ln \left( \frac{Q^2}{Q_0^2}\right)$
dependence is also suggested by analyzing the 
infrared renormalon contribution in $Q_T$ distribution.

\subsection{Testing the universality of 
$\widetilde{W}_{j\bar{k}}^{NP}$}

The CSS resummation formalism suggests the non-perturbative function 
to be universal. Its role is similar to the parton distribution 
function (PDF) in any fixed order perturbative calculation,
and its value needs to be determined by data.
The first attempt to determine such a universal non-perturbative function 
was done by Davies, Webber and Stirling (DWS) \cite{DWS}
in 1985 using Duke and Owens parton distribution function.
In 1994, Ladinsky and Yuan (LY)~\cite{LY} observed that the prediction 
of the DWS set of $\widetilde{W}_{j\bar{k}}^{NP}$ 
largely deviates from the R209 data 
($p+p \ra \mu^+ \mu^- + X$ at $\sqrt{S}=62$\,GeV) using the CTEQ2M PDF.
To incorporate possible $\ln(\tau)$ dependence,
LY postulated
\bea
\widetilde{W}_{j\bar{k}}^{NP}(b,Q,Q_0,x_1,x_2)={\rm exp}\left[
-g_1b^2-g_2b^2\ln \left( {\frac Q{2Q_0}}\right) -g_1g_3b\ln {(100x_1x_2)}
\right] ,
\ena
where $x_1x_2=\tau$.
A ``two-stage fit'' of the R209, CDF-$Z$ ($4\,{\rm pb}^{-1}$ data) 
and E288 ($p+Cu$) data gave 
$g_1=0.11_{-0.03}^{+0.04}~{\rm GeV}^2$, 
$g_2=0.58_{-0.2}^{+0.1}~{\rm GeV}^2$, and
$g_3=-1.5_{-0.1}^{+0.1}~{\rm GeV}^{-1}$, for $Q_0=1.6~{\rm GeV}$
and
$b_{max}=0.5~{\rm GeV}^{-1}$.
Unfortunately, a fortran code error in calculating the  
parton densities inside the neutron
lead to a wrong $g_3$ value. 
(An erratum with corrected values will be submitted after 
the completion of our revised analysis.)
Currently, Brock, Ladinsky, Landry, and Yuan~\cite{BLLY}
are revisiting this problem
using the R209 ($p+p$), 
CDF-$Z$ ($p + \bar p$ with $4\,{\rm pb}^{-1}$ data), 
E288 ($p+Cu$),
and E605 ($p+Cu$) data, with CTEQ3M PDF.
The preliminary results show that, for $Q_0=1.6$\,GeV and 
$b_{max}=1/(2\,{\rm GeV})$, 
two forms for $\widetilde{W}_{j\bar{k}}^{NP}(b,Q,Q_0,x_1,x_2)$
give good fits ($\chi^2/dof \simeq 1.4$):
(i) $g_1=0.24$, $g_2=0.34$ and $g_3=0.0$
(DWS form, pure Gaussian form in $b$-space, without $x$ dependence),
and
(ii) $g_1=0.15$, $g_2=0.48$ and $g_3=-0.58$
(LY form, \\
with a linear $b$ term and $x$ dependence).
We are in the process of determining the uncertainties of these
fitted parameters $g_i$.

\subsection{Run-1B $W/Z$ data at the Tevatron}
            
\begin{figure}
\centerline{
\psfig{figure=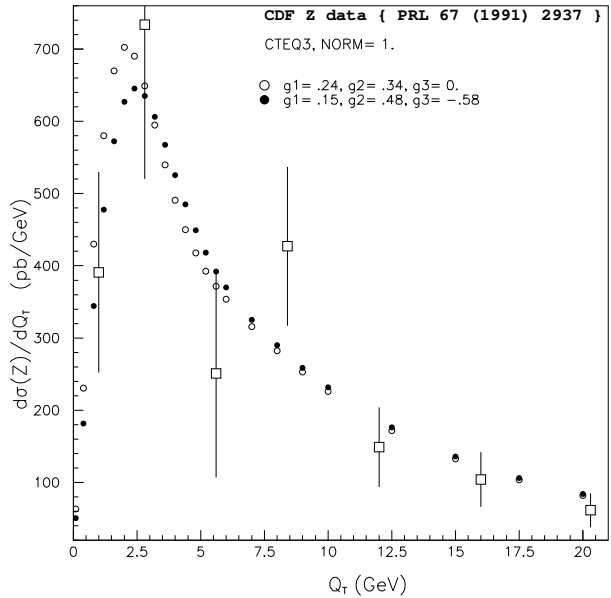,height=3.5in}
}
\caption{Comparison of $4\,{\rm pb}^{-1}$ CDF-$Z$ data with two 
different theory model predictions.
\label{fig:cdfZ}}
\end{figure}

The Run-1B $W/Z$ data at the Tevatron can be useful  
to test the {universality} and the $x$ dependence of 
the non-perturbative function
$\widetilde{W}_{j\bar{k}}^{NP}(b,Q,Q_0,x_1,x_2)$.
In Fig. \ref{fig:cdfZ}, we show the prediction of the two different
global fits (2-parameter and 3-parameter fits)
obtained in the previous section using the CTEQ3M PDF.
(The CTEQ4M PDF gives the similar results.)

We note that for $Q_T > 10$\,GeV, 
the non-perturbative function has little effect on the $Q_T$ distribution 
although in principle it affects the whole $Q_T$ range (up to $Q$).
This is also clearly shown in the figures of Ref. \cite{wres}. 
(Because of the limited space of this article, I will not reproduce 
here the figures that can be found in Ref. \cite{wres}.)
To study the resolution power of the Tevatron 
Run-1B $Z$ data on determining the non-perturbative function,
we have performed a ``toy global fit'' as follows.
First, we generate a set of ``fake Run-1B $Z$ data'' 
(assuming 5,000 reconstructed $Z$ bosons) using the original 
LY fit ($g_1=0.11$, $g_2=0.58$ and $g_3=-1.5$).
Then, we combine this set of ``fake'' data with the low energy 
Drell-Yan data as listed above to perform a global fit.
Using the 3-parameter form (the LY form), we get back the $g_1$ and $g_2$ 
values from the fit but the $g_3$ value is smaller by a factor of 2. 
The best fit gives
$g_1=0.11$, $g_2=0.57$ and $g_3=-0.88$.
(This amounts to shift the prediction on the mass and the width of 
the $W$ boson by about 5\,MeV and 10\,MeV, respectively \cite{new}.)
With this large sample of $Z$ data, it can be clearly illustrated that 
a single parameter without $Q$ dependence (i.e. the parameter $g_1$ alone) 
cannot yield a good global fit.

\section{$W/Z$ Production and Decay at the 1.8 TeV Tevatron}

In Ref. \cite{wres}, we have presented the results of a detailed study on 
the distributions of the decay leptons from the $W/Z$ boson produced at 
the Tevatron. Here, we shall only present a few of the 
most interesting results.

\subsection{``Matching'' from low to high $Q_T$}

First of all, let us examine 
the distribution of the transverse momentum of $W$ boson 
produced at the 1.8\,TeV Tevatron.
It is obvious that the 
CSS formalism gives automatic matching in the $Q_T$ distribution 
if $A,B,C$ and $Y$ functions were known to all orders in $\alpha_s$.
However, in practice we only calculate those functions to some fixed 
order in $\alpha_s$, therefore, some ``switching'' procedure 
(to switch from the total resummed result to the fixed-order perturbative
result) 
should be applied to have a continuous distribution that would match the 
fixed-order result in the large transverse momentum $Q_T$ 
(of the order $Q$) where 
$\ln(Q^2/Q^2_T)$ is small and the resummed result is less accurate.
Our matching prescription is to switch from the resummed prediction to the 
fixed-order perturbative calculation as they cross around $Q_T \sim Q$.
This switching procedure is done for any given $Q$ (invariant mass) 
and $y$ (rapidity) of the $W/Z$ boson.
We note that this procedure is different from that proposed in 
Ref. \cite{arnold}.
As shown in the figures of Ref. \cite{wres}, 
including a higher order $Y$ term, as expected, improves 
the ``matching'' between the resummed and the fixed-order results.
At $O(\alpha_s)$, that crossing occurs at about 50\,GeV, while at
$O(\alpha_s^2)$, it occurs at about 70\,GeV for $W$ boson 
production.

\subsection{Total cross section}

It was shown by an analytical calculation in Ref.~\cite{wres} that 
if ``matching'' (switching) is chosen to be at 
$Q_T=Q$, then the CSS resummed rate is the same as the NLO rate.
However, in that case, the $Q_T$ distribution is not smoothly continuous 
for $Q_T$ close to $Q$. Since
the region of $Q_T > 50$\,GeV only contributes to the total rate 
by less than 2\%, and the CSS resummed formalism includes also some
even higher order contributions beyond NLO, 
we regard the difference between the 
resummed rate and the NLO rate as a gauge of the uncalculated
higher order contribution which is shown to be small at the Tevatron.
\vspace{5mm}
\begin{center}
{Total cross sections of $p {\bar p} \rightarrow (W^+ {\rm or}~ Z)
X$ at the Tevatron, calculated in different
prescriptions, in units of nb.}
\\
{\normalsize
\begin{tabular}{l|c|cc|c|c|c|r}
$V$ & $E_{cm}$ & \multicolumn{2}{c|}{Fixed Order} 
& Res. (1,1,1) & Res. (2,1,1) & Res. (2,1,2)
& Experiment \\ 
& (TeV) & ${\cal O}(\alpha _S^0)$ & ${{\cal O} (\alpha _S)}$ & $\oplus$
Pert. ${\cal O}(\alpha _S )$ & $\oplus$ Pert. ${\cal O}(\alpha _S )$ & $%
\oplus$ Pert. ${\cal O}(\alpha _S^2)$ &  \\ \hline
W$^{+}$ & 1.8 & 8.81 & 11.1 & 11.3 & 11.3 & 11.4 & 11.5 $\pm $ 0.7 \\ 
W$^{+}$ & 2.0 & 9.71 & 12.5 & 12.6 & 12.6 & 12.7 &  \\ 
Z$^0$ & 1.8 & 5.23 & 6.69 & 6.79 & 6.79 & 6.82 & 6.86 $\pm $ 0.36 \\ 
Z$^0$ & 2.0 & 6.11 & 7.47 & 7.52 & 7.52 & 7.57 & 
\end{tabular}
}
\end{center}
\vspace{5mm}
In the above Table,
the values of the strong coupling constants used with the CTEQ4L and CTEQ4M 
PDF's are $\alpha _S^{(1)}(M_{Z}) = 0.132$ and 
$\alpha _S^{(2)}(M_{Z}) = 0.116$, respectively.
(Res. (2,1,2) denotes the result with 
$A$ and $B$ calculated to $\alpha_S^2$ order, $C$ to $\alpha_S$, and
$Y$ to $\alpha_S^2$ order, {\it etc.})
We note that the total rate is
not sensitive to either non-perturbative function or matching prescription.

\subsection{Lepton transverse momentum}

\begin{figure}
\centerline{
\psfig{figure=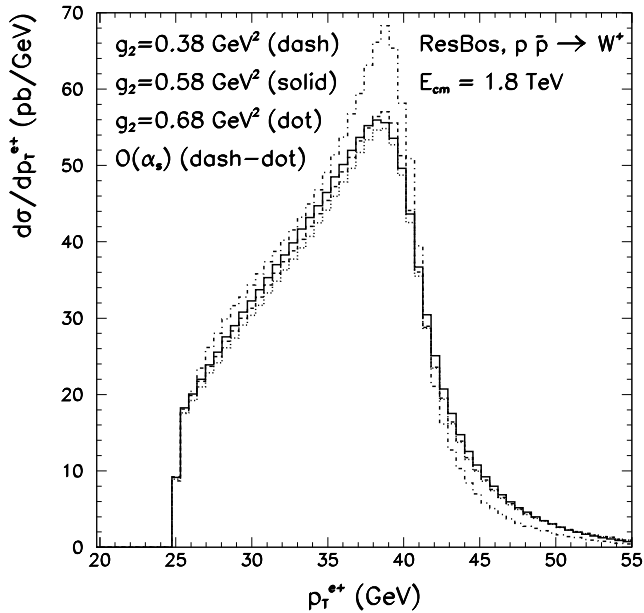,height=3.5in}
}
\caption{
NLO (dashed) and resummed ${\cal O}(\alpha _S)$ (solid) calculations 
Using the LY fit 
with cuts ($Q_T<30~{\rm GeV}, p_T^{e^+,\nu}>25~{\rm GeV}$).
\label{fig:pte}}
\end{figure}

At the Run-2 of Tevatron, the number of interactions per crossing
increases, and the transverse mass measurement becomes less accurate.
It is useful to also measure the inclusive lepton $P_T^{e^{+}}$ spectrum.
In Fig. \ref{fig:pte}, we show that the difference between the 
resummed and the NLO results is much larger than the dependence on the 
non-perturbative function in the resummed calculation (although
its dependence is not negligible).

\subsection{Lepton charge asymmetry}

The distribution of the rapidity of the charged lepton 
from $W^\pm$ boson decay provides useful information to determine
the ratio of the up and down quark parton distribution functions.
It was shown in Ref. \cite{wres} that 
without imposing the kinematic cuts, the prediction of the
resummed calculation is the same as that of the NLO.
This is obvious because an integration over the entire $Q_T$ space 
in the CSS formalism reproduces the NLO rate as discussed above.
However, with kinematic cuts, the rapidity distribution 
of the charged lepton or the $W$ boson is different 
between the resummed and the NLO calculations.
Furthermore, for this distribution, the dependence on the non-perturbative function is 
negligible.

The resummed calculation also predicts different 
correlations between the two decay leptons from $Z$ boson.
A couple of examples are the distribution showing the
balance in transverse momentum
$\Delta {p}_T=|{\vec p}_T^{\,\ell_1}|-|{\vec p}_T^{\,{\bar \ell_2}}|$,
and
the angular correlation ${z=-\vec p}_T^{\,\ell _1}\cdot 
{\vec p}_T^{\,{\bar \ell_2}}/[\max 
(p_T^{\ell_1},p_T^{{\bar \ell_2}})]^2$.
We refer the reader to Ref. \cite{wres} for more details.

\section{Di-photon rates and distributions}

It is straightforward to extend the above CSS formalism to 
calculate the distribution
of photons from $h_1 h_2 \ra \gamma \gamma X$.
In Ref. \cite{diphoton}, we have 
included the full content of NLO contributions from $q {\bar q}$ and $q g$
subprocesses. The accuracy of the resummed result is similar to that 
for the Drell-Yan and $W/Z$ productions.
In addition, we also included part of the higher order contribution 
from the $gg$ process whose lowest order contribution comes from 
one-loop box diagrams. This was done by including $A^(2)$ term in the 
Sudakov factor to resum higher order large logs due to initial state
radiation, and the NLO contribution,
of ${\cal O}(\alpha_{em}^2\alpha_s^3)$,
to the hard scattering subprocess is added in an approximate fashion.  

\begin{figure}
\centerline{
\psfig{figure=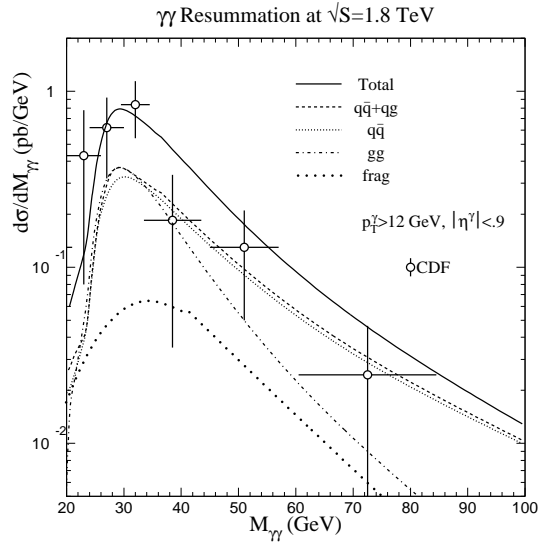,height=3.0in}
}
\caption{The predicted distribution for the invariant mass of the photon
pair $M_{\gamma\gamma}$ from the resummed calculation compared to the
CDF data, with the CDF cuts imposed in the calculation.
\label{fig:CDF0}}
\end{figure}

We found that the $gg \ra \gamma \gamma X$ rate is not small 
at the Tevatron.
Including part of the NLO $gg \ra \gamma\gamma$ 
approximately doubles the
LO $gg$ box contribution to the cross section.
This is clearly seen in the distribution of the invariant mass 
of the photon pair, as shown in Fig. \ref{fig:CDF0}.
The other distributions, such as the distribution of the 
transverse momentum of the photon pair $Q_{T}$, compared with CDF and \D0 
data, can be found in Ref. \cite{diphoton}.
We note that to accurately predict the production rates of the di-photon
pairs at the CERN Large Hadron Collider
(LHC) requires a better calculation of the $gg$ fusion rate.
Namely, a full NLO result is needed to predict the distribution 
of the di-photon pair with large $Q_T$ or small $\Delta \phi$.

In Ref. \cite{diphoton}, we also presented our results
for the di-photon production rate and distributions at the
fixed-target experiment (E706 at Fermilab):
$p+Be\to\gamma\gamma + X$ at $\protect{\sqrt{S}}$=31.5\,GeV.
In this case, we found that the $gg$ rate is small,
and the distribution of the 
transverse momentum of the photon pair $Q_{T}$
is dominated by the non-perturbative contribution.
Therefore, this data can be useful for determining the 
non-perturbative function associated with the $gg$ initiated 
processes in the CSS formalism.
We refer the readers to Ref. \cite{diphoton} for more details.

\section{Conclusion}

In conclusion, the 
effects of QCD gluon resummation are important in many precision 
measurements.
A Monte Carlo package {\underline {ResBos}}
({ ({\underline{Res}}ummed 
{\underline {Bos}}on Production and Decay)})
is available~\cite{csaba} for studying the effects of 
gluon resummation (with NLO hard part corrections) 
in hadron collisions.

Before closing, I would like to comment on the $Q_T$ space resummation 
formalism proposed in Ref. \cite{ev}, and compare that with the
$b$ space resummation formalism (CSS formalism) discussed in this article.
Despite of its imperfectness in the theory structure, in practice, 
the $Q_T$ space resummation formalism may still prove to be useful because 
it does not require a ``switching'' in the large $Q_T$ region and 
its calculation takes much less CPU time due to the fact that 
there is no need to perform a Fourier transformation from the $b$ space to 
the $Q_T$ space. 
This is obvious for $Q_T$ above 10 GeV where the 
non-perturbative contribution is not important, as discussed above.
However, in the small $Q_T$ region, it is not obvious that both formalisms
will give the same prediction of the $Q_T$ distribution of the $W^\pm$
boson after the non-perturbative part is fit by the $Q_T$ of the $Z$ boson
using the Tevatron Run-1B and Run-2 data.
Nevertheless, if one is not interested in testing the universality
of the non-perturbative function, as suggested by the CSS formalism, then
it seems likely that it is possible to choose a proper form of the 
non-perturbative function in the $Q_T$ space resummation formalism 
to reproduce (within the experimental uncertainties)
the prediction of the $b$ space formalism when considering 
the Tevatron $W/Z$ data alone. If one is interested in the universality 
property of the formalism\footnote{
Needless to say, this postulation has to be further tested by data.}
so that one can use the same values of the
non-perturbative function to predict future data at different hadron
colliders (such as the LHC), then it becomes unclear whether these two
formalisms will always give the same physics prediction.

\section*{Acknowledgments}

This was the first time I participated the Moriond meeting, and 
I enjoyed it very much. I would like to thank the organizers for 
inviting me to present such a short review talk. 
I would also like to thank my collaborators: 
 C. Bal\'azs, E. Berger, R. Brock, G. Ladinsky, F. Landry,
 S. Mrenna, and J.W. Qiu for the fruitful collaborations. From them and my 
CTEQ colleagues, I have learned a great deal of QCD physics.
Finally, I thank K. Ellis for discussion on the $Q_T$ space formalism.
This work  was supported in part by the NSF grant
No. PHY-9507683.


\section*{References}
\begin{thebibliography}{99}

\bibitem{DDT}  Y.I. Dokshitser, D.I. D'Yakonov, S.I. Troyan, Phys. Lett. 
{\bf B79} (1978) 269.

\bibitem{Parisi}  G. Parisi, R. Petronzio, Nucl. Phys. {\bf B154} (1979)
427.

\bibitem{CSS}  
 J. Collins, D. Soper, Nucl. Phys. {\bf B193} (1981) 381; 
Erratum {\bf B213} (1983) 545; {\bf B197} (1982) 446; \\
J. Collins, D. Soper, G. Sterman, Nucl. Phys. {\bf B250}
(1985) 199.

\bibitem{ev}
R.K. Ellis and S. Veseli, Nucl. Phys. {\bf B511} (1998) 649.

\bibitem{wres}
C. Bal\'azs and C.--P. Yuan, Phys. Rev.  {\bf D56} (1997) 5558,
and the references therein. 

\bibitem{sterman}
G.P. Korchemsky and G. Sterman, Nucl. Phys. {\bf B437} (1995) 415.

\bibitem{DWS}  C. Davies, B. Webber, W. Stirling, 
Nucl. Phys. {\bf B256} (1985) 413.

\bibitem{LY}  G.A. Ladinsky, C.--P. Yuan, Phys. Rev. {\bf D50}
(1994) 4239.

\bibitem{BLLY}
R. Brock, G. Ladinsky, F. Landry, and C.--P. Yuan,
in preparation.

\bibitem{new}
E. Flattum, private communication;\\
B. Ashmanskas, private communication. 

\bibitem{arnold}
 P.B. Arnold, R.P. Kauffman, Nucl. Phys. {\bf B349} (1991) 381.

\bibitem{diphoton}
C. Bal\'azs, E.L. Berger, S. Mrenna, and~~C.--P. Yuan,
Phys. Rev. {\bf D57} (1998) 6934, and the references therein.

\bibitem{csaba}
It can be obtained from http://www.pa.msu.edu/$\sim$balazs/ResBos/. 

\end{thebibliography}
\end{document}